# NETWORK TRAFFIC ADAPTATION FOR CLOUD GAMES

Richard Ewelle Ewelle[1], Yannick Francillette[1,] Ghulam Mahdi[1]
And Abdelkader Gouaïch [1]
[1]LIRMM, University of Montpellier, France, CNRS

***ABSTRACT***

*With the arrival of cloud technology, game accessibility and ubiquity have a bright future; Games can be hosted in a centralize server and accessed through the Internet by a thin client on a wide variety of devices with modest capabilities: cloud gaming. However, current cloud gaming systems have very strong requirements in terms of network resources, thus reducing the accessibility and ubiquity of cloud games, because devices with little bandwidth and people located in area with limited and unstable network connectivity, cannot take advantage of these cloud services.*

*In this paper we present an adaptation technique inspired by the level of detail (LoD) approach in 3D graphics. It delivers multiple platform accessibility and network adaptability, while improving user's quality of experience (QoE) by reducing the impact of poor and unstable network parameters (delay, packet loss, jitter) on game interactivity. We validate our approach using a prototype game in a controlled environment and characterize the user QoE in a pilot experiment. The results show that the proposed framework provides a significant QoE enhancement.*

## 1. INTRODUCTION

Video games have seen tremendous popularity and growth in the last few years. At the same time, there has been significant interest in cloud computing, which is seen by many as the next big paradigm in information technology. The benefits of moving to the cloud have already been discussed in the literature, e.g. [3]. They include scalability, accessibility and computational capability. The combination of video games and cloud computing will immensely increase the use of video games, by enabling them to reach geographically separated storage or data resource with even cross-continental-networks. With this configuration, the performance degradation of the underlined networks will surely affect the cloud game performance and the players' quality of experience (QoE). QoE can be defined as the perceived quality of a service or an application by a user.

Cloud gaming, also called "gaming on demand", is the type of online gaming that works on the same principle as does video on demand which is another data-intensive application. It allows direct and on-demand streaming of video games on a computing device through the use of a thin client. Here, the actual game is stored on the operator's server and it is directly streamed to the device. Due to these potential advantages, many companies like Onlive [19], G-Cluster [11], StreamMyGame[22], Gaikai and T5-Labs[23] are offering cloud gaming services.

However, these data-intensive cloud services have very strong requirements in terms of network resources. In fact in order to ensure the delivery of quality of service (QoS) for the bulk data transfer generated by these applications, a certain amount of network resources should be available. These constraints reduce the accessibility and ubiquity of such services, because devices with little bandwidth capabilities and people located in area with limited network connectivity, cannot take advantage of these cloud services. In addition, the number of connected users increases every day, creating more processing demand, and network congestion, thus limiting the use of cloud services. In the context of cloud gaming, to preserve game accessibility, we need not only to weaken these network constraints, but also to maintain the quality of the game.

The main contribution of this paper can be summarized as follows: We propose an organization based network traffic adaptation with the cloud gaming paradigm for accessibility, adaptability and player QoE maintenance. We focus in our approach on the organization of game entities as a means to express different priorities among entities depending on their importance in the game. Whenever the required network resources for the game state synchronisation is not meet by the actual network configuration, our framework ensures that the entities with the high priority regarding the communication are given more network resources than the ones with low priority, thus enables to meet the limited network constraints on poor bandwidth settings. Our approach also makes it possible to adapt the game to unstable network parameters with less impact on the player QoE.

Our approach focuses on the interaction between video games and the cloud platform. The general question guiding our study is: ***How to build and run video games efficiently over the cloud by optimizing the interaction between the client and the server for an increased player experience and game accessibility for poor and variables bandwidth networks?***

User experience in games and more generally in interactive entertainment systems has been a real focus in the research community. In cloud gaming, many factors can affect user quality of experience. Following the study in [24], quantitative measurements of user experience mainly depends on subjective factors like response time and graphics quality or received video quality. The game response time is the total delay from the user control command occurring, to the corresponding graphic or video frame displaying on the device's screen. It is influenced by network parameters like bandwidth, delay, packet loss and jitter. While the received game video quality is influenced by the image quality in each frame and the smoothness of all the frames which are affected by video settings like resolution, data rate, frame rate, codec, etc.

All these factors affect the game response time and game video quality in a complex manner. Thus, for this paper, we focus on optimizing the game system in other to minimize the effect of poor network parameters for a better game interactivity and an increased user experience.

The rest of the paper is organized as follows: Section 2 discusses the background of the work. Section 3 presents the related work which is followed by our proposed adaptation approach in section 4. Section 5 describes the experimental framework and the game used to validate our approach; we then analyse the results of our pilot experiment with player evaluation in terms of QoE and finally, we conclude this paper in section 6 by presenting conclusion and some future directions for this work.

## 2. BACKGROUND

### 2.1. Cloud computing

While cloud computing is currently a term without a single consensus meaning in the marketplace, it describes a broad movement toward the use of wide area networks, such as Internet to enable interaction between IT service providers of many types and consumers. According to [12], a cloud is a hardware and software infrastructure which is able to provide services at any traditional IT layer: software, platform, and infrastructure, with minimal deployment time and reduced costs. Along with this new paradigm, various business models are developed, which can be described by terminology of"X as a service (XaaS)"4, where X could be software, hardware, data storage, and etc. Examples includes Amazon's EC2 and S31, Google App Engine13, and Windows Azure16 which provide users with scalable resources in the pay as you use fashion at relatively low price.

Hosting an application in the cloud allows the application provider to continuously update the application without issuing and shipping new installation disks. The user will always get the latest version of the application. Cloud Computing offers to application providers the possibility to deploy a product as a SaaS (Software as a service) and to scale it on demand, without building or provisioning a data center themselves. With respect to these considerations, cloud gaming is then the offering of game applications through the cloud to players. The SaaS application here is the video game.

## 2.2. Cloud gaming

Cloud gaming can be defined as just executing games in a server instead of users' devices. In a client-server setup with the cloud gaming paradigms, two trends exist: Video streaming and game state synchronisation.

### 2.2.1. Video streaming

The most used solution in industrials cloud gaming systems are based on video streaming [19],[11],[22],[23]. In this configuration, only the server maintains a representation of the game model; the client simply gathers inputs from the user and sends them to the server; and receives video from the server and displays it to the user's screen.

When launched the client contacts the server to establish the connection and start the game. As explained in [7], usually there are two connections: one connection to send the user's controller input to the server and one connection to receive the video from the server. The figure 1 shows this game loop.

The client loop: the video task and the input task:

- The client polls for input events from the player. When an event is received it is written into a packet and sent to the server.
- The client receives packets from the server; these packets are used to decode the video stream. When a complete frame has been decoded, it is displayed in the application's window.

A typical server game loop also runs two tasks:

- An input task that receives inputs from the client application and passes those inputs to the game model.
- A video streaming task that captures frames from the game, encodes them into a video, and then sends the video output to the client.

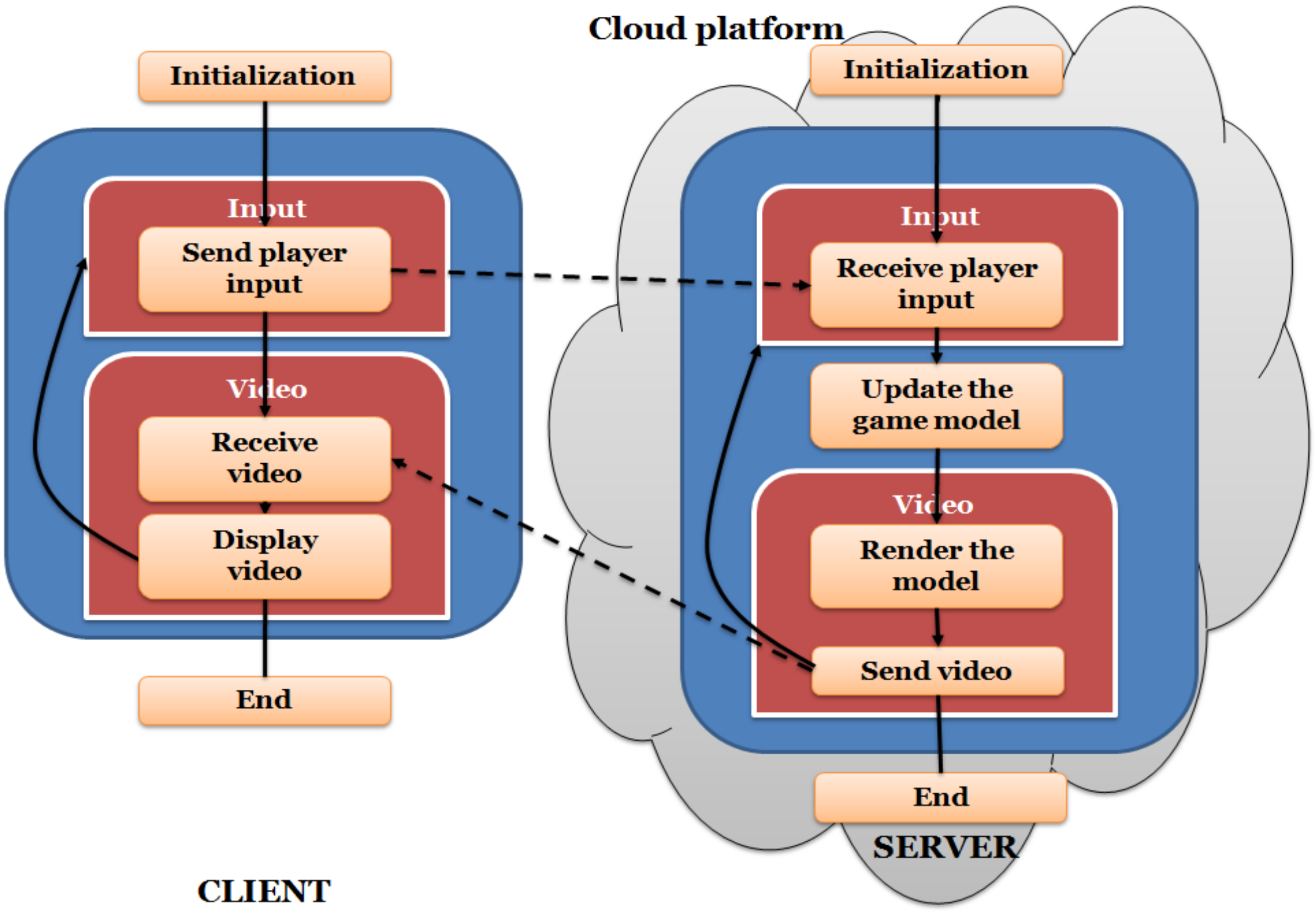

Figure 1: Cloud game loop: Video streaming

### 2.2.2. Game state synchronisation

The client and the server have a representation of the game model. The client side periodically synchronizes his local game model with the server side remote game model which is the central one. This is done by receiving update messages about state change in the game model. This game update loop is presented in the figure 2. The game usually starts with an initialization stage where all the objects and entities of the game model are initialized. After this phase, the game loops between the input phase and the update phase. The major difference is that, with this configuration, the video rendering is done on the client side, and the server just sends state updates in smaller packets.

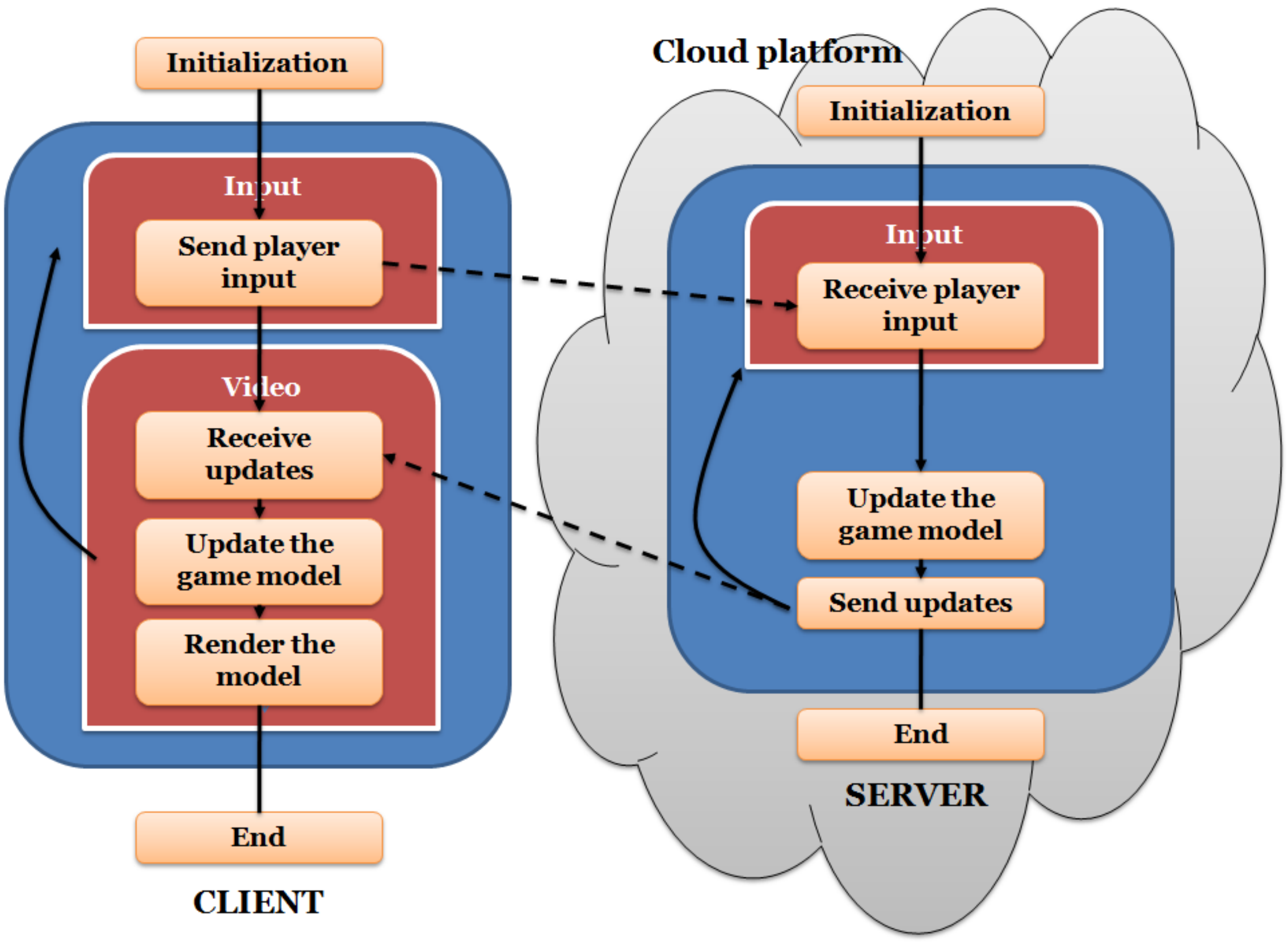

Figure 2: Cloud Game loop: State synchronisation

In this paper, we will consider the game state synchronisation approach as an alternative to video streaming in cloud gaming because it does not have strong requirements on network resources.

## 3. Related Work

In the cloud gaming paradigm, to improve the interactivity performance of a game architecture, two main causes for delays have to be analysed: network latencies and video processing costs. Several research works have already brought contributions to the optimization of the video processing mechanism, in our study we are focusing on network optimization.

For game state synchronization in client-server games, several research studies have already contributed to the development of efficient synchronization schemes. In particular, packet compression [2] tries to speed up transmissions by reducing bandwidth requirements. Indeed, minimizing the number of bits needed to represent game information is a proficient method to diminish the traffic present in the network. Aggregation is another technique attempting to reduce the overhead associated with each transmission therefore limiting the bandwidth required by the application. Specifically, before being transmitted, packets are merged in larger ones thus reducing the overhead. Both schemes, however, pay the latency benefits achieved with an increment in computational costs. Information compressed and aggregated, needs to be recovered with decompressing and disaggregating algorithms at the receiving end, thus increasing the time required to process each single event. Moreover, aggregation can origin further waste of time if a transmission is delayed while waiting for having available other events to aggregate. The packet compression technique is often used in video game streaming.

To reduce both the traffic load in the network and the computational cost to process each game event, Interest Management techniques have been very beneficial [17]. In some game scenario, events generated are relevant only for a small fraction of the users. Therefore, implementing an area-of-interest scheme for filtering events, as well as a multi-cast protocol, could be put in good use to match every packet with the nodes that really need to receive it and, consequently, to reduce both the traffic and processing burden at each node [9]. Games having interest areas occupying a significant portion of the global virtual environment could hence be further delayed if Interest Management schemes would be implemented.

In order to be less dependent on the real responsiveness of the network, optimistic algorithms for synchronizing game state at servers can be implemented to avoid delay perception at destination. In case of lousy interactivity between client and server, in fact, an optimistic approach executes events before really knowing if ordering would require to process other on the way events first. Game instances are thus processed without wasting any time in waiting for other eventually coming packets. On the other hand, this performance gain is paid with some occurrence of temporary consistency loss.

Dead reckoning is another method that can help to minimize the effects of latency, but it can also introduce some temporary incoherence between the factual game state and the assumed one at the server [20]. In fact, attempting to limit the bandwidth required by the application, this scheme utilizes a reduced frequency in sending update packets while compensating the lack of information with prediction techniques. Obviously, predicted movements and actions are not always trustful. These eventual restoring actions further impact on interactivity and playability of the game.

In the framework Games@Large[14], A. Jurgelionis et al. propose a distributed gaming platform for cross-platform video game delivery. The server executes the game, captures the graphic commands and sends them to the client. The commands are rendered on the client device allowing the full game experience. For client devices that do not offer hardware accelerated graphics rendering, the game output is locally rendered at the server and streamed as video to the client. The advantage of this approach is that the architecture is transparent to legacy code and allows all type of games to be streamed. For devices with rendering capabilities, it discharges the server from rendering and encoding the game output, enhancing the server performance, it also reduce the amount of data to be transmitted since only graphics commands are sent to the client, thus reduces the game latency. For users with low end devices, the latency of video encoding and video transmission remain the same as in other cloud gaming architecture like Onlive.

These approaches propose enhancements that can improve the performance of a real-time networked game by reducing the traffic load between the client and the server. Nonetheless, this traffic reduction introduces some computation expenses contributing to the lag or other incoherence in the game. Techniques such as interest management and dead reckoning are widely used to reduce both the network load and the computational cost for a better player quality of experience. None of the analysed techniques has tried to couple the traffic reduction with a scheduling mechanism for an efficient message passing (messages with different importance) between the client and the server. And to the best of our knowledge, we haven't found any work that tries to automatically adapt the traffic generated by the game to the actual network conditions at runtime.

We present here a novel synchronization adaptation technique, specifically designed for efficient event delivery in client-server games with the cloud paradigm.

# 4. ADAPTATION FRAMEWORK

This section introduces the Level of Detail (LoD) principles and presents our proposed approach for client-server game state synchronisation.

## 4.1. Overview

In classical networked game architecture with the cloud gaming paradigm, all the game entities of a scene are updated in a synchronous basis. We aim at optimizing the state synchronization process between the client and the server. We make the assumption that there are different synchronization needs per game entity. Some need a small update frequency and for others a larger one will be enough. For example, background entities need less synchronization than target entities in a shooting game. We therefore need an efficient message passing protocol that takes entity's synchronisation needs in consideration for better performances. That is why we prioritize some updates over others. This enables us to lower the game's communication requirements while maintaining a good update frequency for more important entities.

The approach maintains a bidirectional multi-level quality of service for game entities at runtime. Meaning that the adaptation needs to lessen the games communication requirements when it notices a bottleneck in the network (materialized by an increase in the response time, or packet loss in case of UDP packets) and maintain an optimal communication frequency otherwise.

The proposed framework uses entity organizations. Here each entity belongs to a synchronization group (with a specific communication frequency) and has a specific role within that group. The group assignment for an entity is done according to its significance in the organization. The significance of an entity can be defined in different ways depending of the game designer's choices. For this paper, we use the functional importance of an entity in the game as significance. The overall specifications of this framework are shown in the figure 4. To conceptualise the different communication needs for an entity, we apply a Level of Detail approach.

## 4.2. Level of Detail

The LoD technique has been widely used in 3D graphics and simulations. The basic purpose of the technique is to modulate the complexity of a 3D object representation according to the distance from which it is viewed (or any other criteria) [15]. As introduced by James Clark [6], this technique is meant to manage processing load on graphics pipeline while delivering an acceptable quality of images.

The technique suggests to structure the rendering details of a 3D object in order to optimize its processing quality if the object's visible details are proportional to the distance from which the object is viewed.

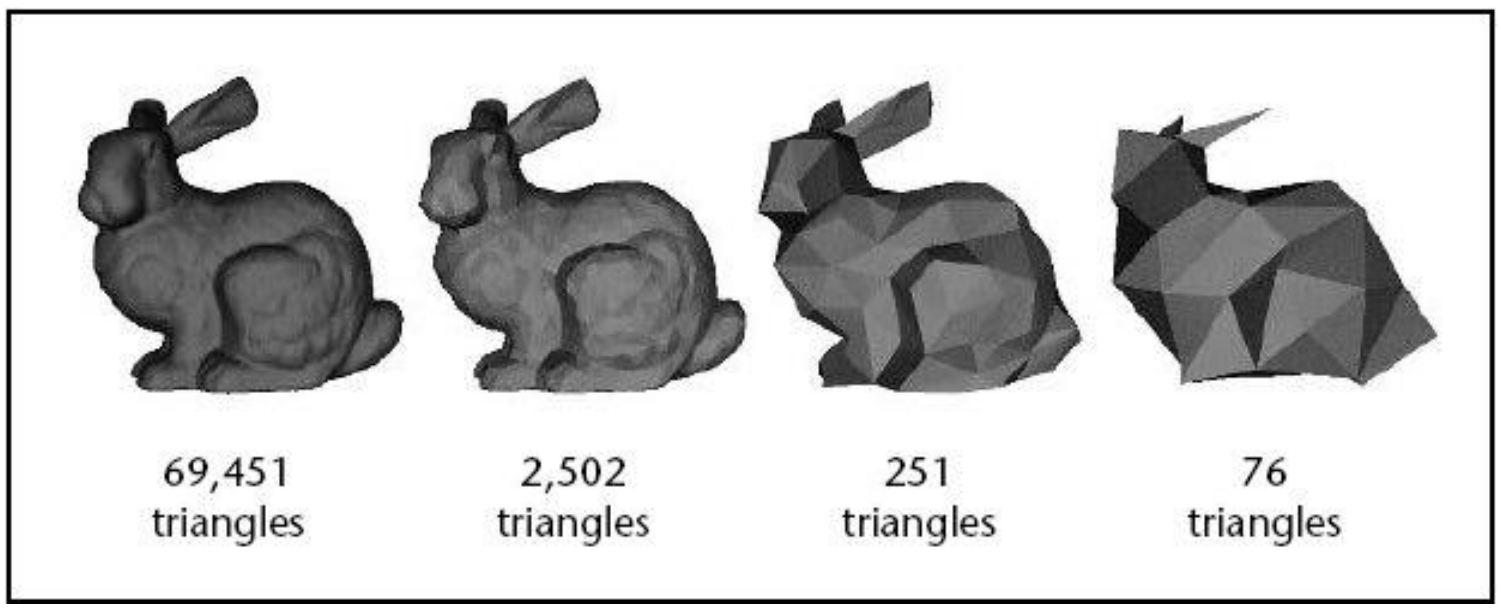

Figure 3: Basic concept of LoD: An object is simplified by different representations through varying number of polygons [15]

The figure 3 presents the rational of Clark: by changing the number of vertices, we see the change in the quality and the visualization of a sphere.

Geometric datasets are usually too large in data size and complex (in terms of time and computational resource demands) so their rendering can become a tedious and time consuming process. The LoD approach suggests different representations of a 3D object model by varying in the details and geometrical complexity. The geometrical complexity of an object is determined by the number of polygons used for his representation. The more complex an object is, the more time consuming its rendering will be.

Although there can be other factors involved in the complexity and resource demand of a graphical model of an object, the relations between polygonal quantity and resource consumption are generally considered as established ones [18]. For example, one can determine the difference in rendering quality by observing the figure 3. This figure also makes it possible to draw a general conclusion on how the number of polygons affects the rendering quality of an image's graphical representational. Once these different representations of a model are on hand (as shown in the figure 3), the LoD technique will suggest their selection at a particular time point based on certain positive selection bias. The latter can be their size, camera distance or any other criteria.

The application of this technique for our work is the ability to have different synchronization frequencies for each game entity, and select one at a particular time based on certain criteria. This way, only more important entities will get the maximum amount of network resources while others get less. These entities can see their communication resources changed when the network situation changes or when their importance in the game changes.

### 4.3. Virtual society of entities

Organizations are used to represent game entities as a virtual society for two main reasons:

- Organizing and representing multiple communication levels for each entity, so that a drastic change in the traffic load steers the use of a different communication level for an entity, therefore a change in the traffic profile of the game.
- Evaluating the relative importance of the entity in real time in order to select the most appropriate communication level.

Our organization model is based on the AGR model [10]. The AGR (Agent, Group, Role) model advocates organization as a primary concept for building large scale and heterogeneous systems. The AGR model does not focus on the internal architecture nor the behaviour of individual agents but it suggests organization as a structural relationship between the collection of agents. The AGR model defines three main concepts as its basis for an organizational structure: agent, as an active and communicating entity; a group is a set of agents which is defined by tagging them under a collection; finally a role defines an agent's functional representation in a group.

The organizational model we use in our approach matches the AGR model as follows:

- **Agent**: An agent represents a game entity involved in the game scene. Each entity has a significance regarding its communication requirements.
- **Role**: The role represents the reason why the entity is in the game. Each entity in the scene has a role and, a role can be shared by several entities. The role can also be used to express entity's communication significance.
- **Group**: A group is a set of entities with the same communication needs. These entities are synchronized at the same frequency. An entity can move from one group to another at any moment according to the observed network settings and his significance in the game.

### 4.4. Communication groups

A communication group is characterized by a communication frequency and a score coefficient (see score coefficient below) threshold, that the entities should satisfy to be assigned to the group as shown in the figure 4.

Each update transmission uses some communication resources for its completion. We describe a score coefficient as an abstract measurement unit for the notion of importance regarding communication resources for weighting entities communication requirements.

An entity's score coefficient is calculated at runtime using a combination of the actual network settings (here for UDP the packet loss percentage) and the entity's significance. This way the entity's importance is proportional to the network congestion at running time. This score coefficient of an entity is computed using the following formula:

***ScoreCoefficient = Significance * CurrentNetworkCongestion***

This generic notion of ScoreCoefficient is a value that defines whether the entity is important at the current time or not. As the significance of an entity is depending to the game rules and the function of the entity, this notion can be exploited in many ways. For the game prototype developed in this paper, the significance is a weight defined on each role by the game designer through a configuration file. In case of congestion, entities are reassigned to groups.

### 4.5. Group assignment

As illustrated in figure 4, the server side of the game engine maintains a collection of communication groups. When the game notices a drastic change in network congestion, the score coefficient of each entity is recalculated and entities are reassigned to communication groups if needed.

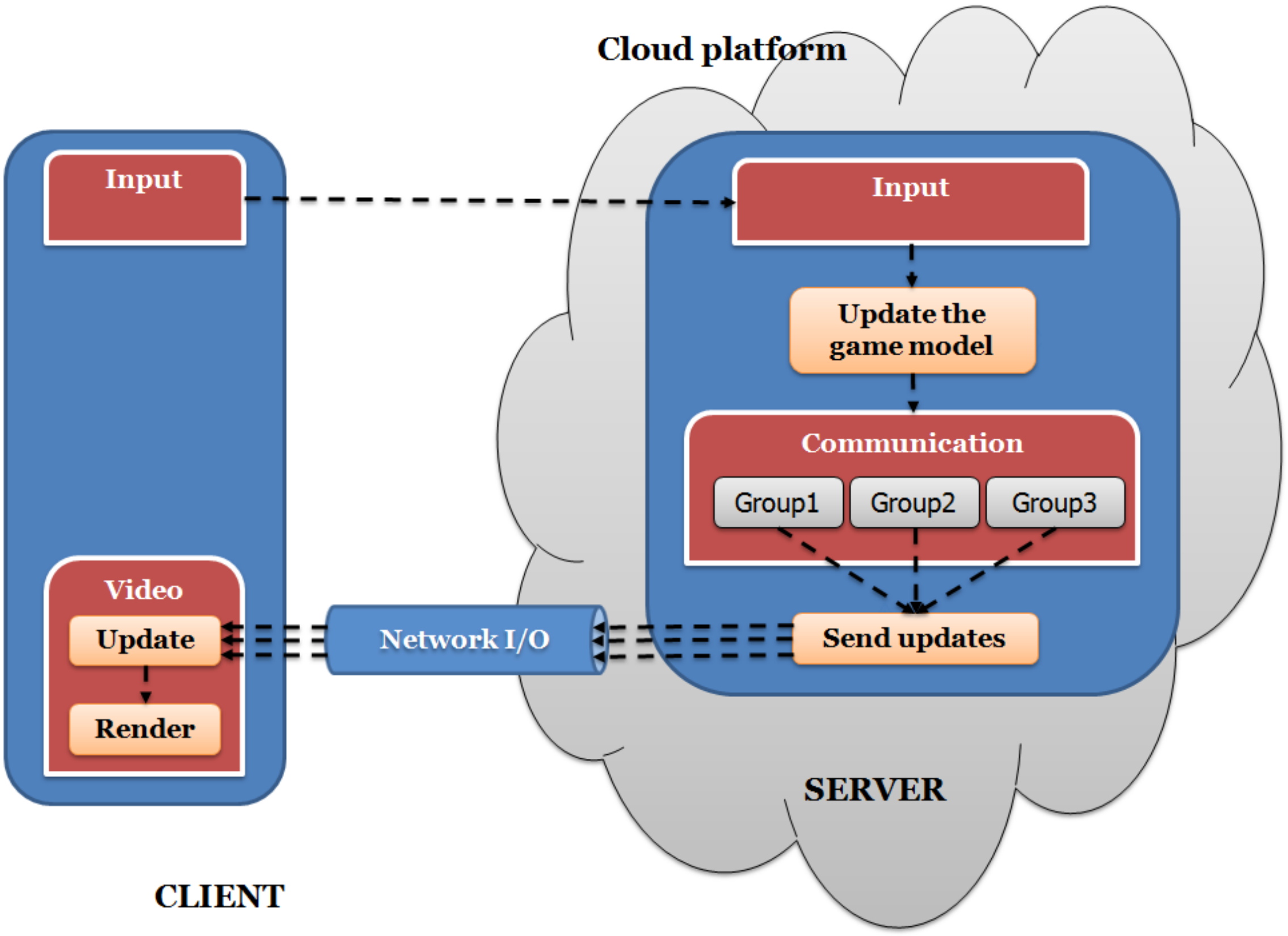

Figure 4: LoD framework specifications

### 4.5.1. Assignment policy

An entity of reference (ER) is used to trigger the group reassignment process for all the entities. The reassignment process is ordered only if the newly computed expected group for the ER is different from its current group as explained in the figure 5.

As general rule for getting the expected group for an entity, these conditions must be respected:

- **C1**: The entity's score coefficient is lower than the group's score coefficient's threshold.
- **C2**: The group is the one with the lowest score coefficient threshold among the groups respecting the condition **C1**.

This reassignment process is attempted every 5 seconds during the entire game session.

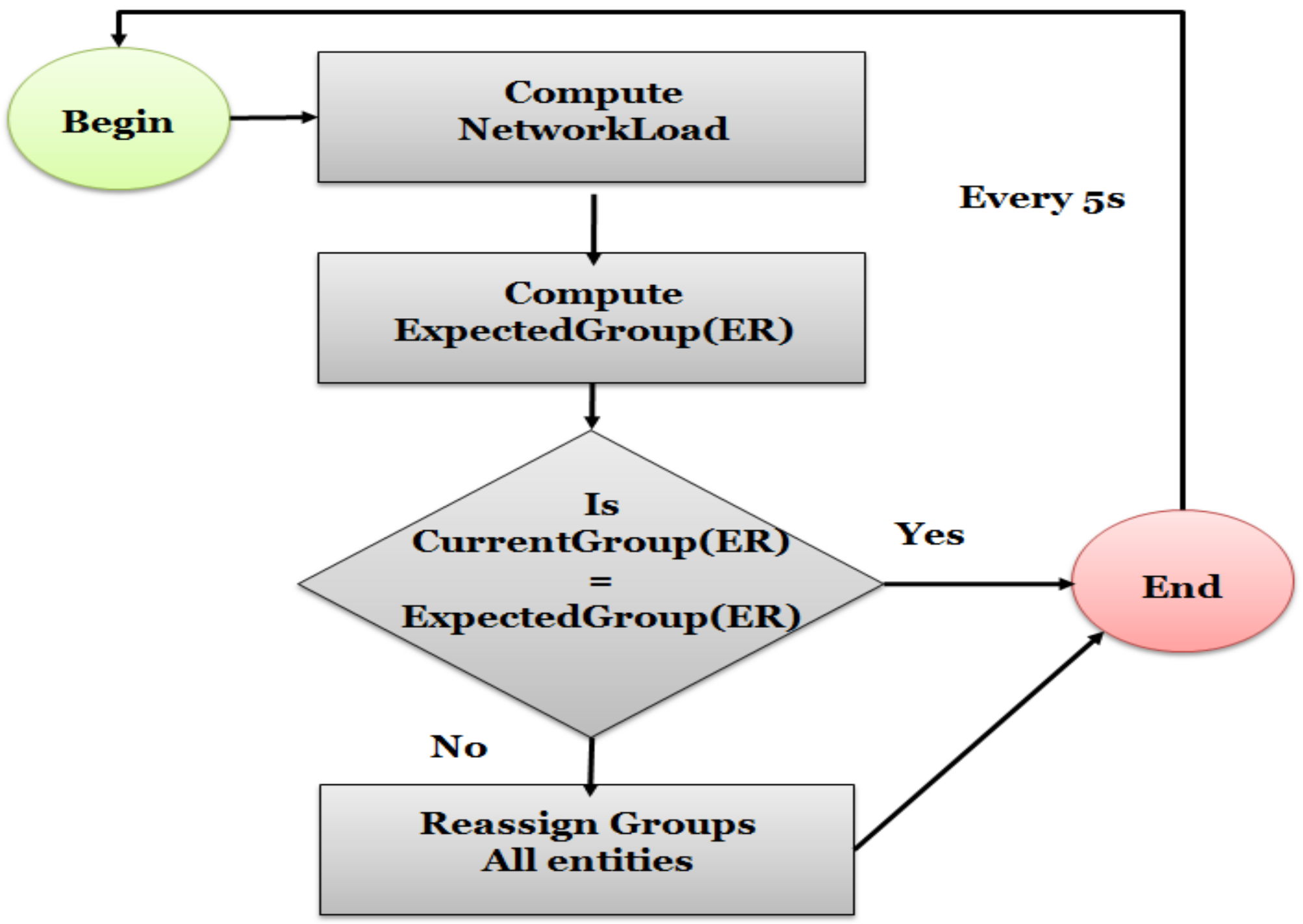

Figure 5: Group assignment policy

The figure 6 shows the group reassignment algorithm.

**<u>Algorithm</u>**: Group assignment
**/* Loop for all entities in the game scene** ***/**
**for** *entity in entities* **do**
    **/* Compute the expected group of the entity** ***/**
    scoreCoefficient = significance(entity)*currentNetworkCongestion;
    expectedGroup = getExpectedGroup(entity, scoreCoefficient);
    **/* Add the entity in the appropriate group** ***/**
    **if** *currentGroup != expectedGroup* **then**
        expectedGroup.add(entity);
        currentGroup.remove(entity)
    **else**
        donothing()
    **end**
**end**

Figure 6: Assignment algorithm

### 4.5.2. Network congestion computing

Our proposition is a generic framework that uses as adaptation metric a score coefficient calculated at runtime. The current congestion of the network is the main parameter for the calculation of this score, and represents the network overhead created by the traffic load at running time. Because our message passing is done using the UDP protocol, as a representative of network congestion, we monitor the packet loss in the network since in case of congestion some packets will be lost. Knowing that with UPD, there is no guaranty on the reception of the packets.

To monitor the packet loss at runtime, the server periodically sends a number of monitoring packets (100 packets, one every 50 ms) to the client and simply counts the number of responses it receives back. Each missing response is marked as packet loss. Thus in case of network congestion, the amount of packet loss will increase, and therefore our adaptation scheme will trigger the reassignment of communication groups for the entities changing the communication profile of the game.

## 4.6. Example

To illustrate how our approach works, here is a simplified version of a shooting game, ”My Duck Hunt”, developed for evaluation purpose. You will find a more complete description in the subsection 5.1. Suppose that we have 3 types of entities: clouds, ducks and a reticle. The player controls the reticle. Let’s say the objective is to point the reticle on the ducks and shoot to kill them. These entities have different functional importance in the game, therefore different significances as defined in the subsection 4.4. Here the significance is a weight value associated with each entity:

- **Cloud**: represents clouds that are moving in the background. Less important; weight: 1.5.
- **Duck**: represents ducks that are flying in the game scene. Ducks are the targets. Medium importance; weight: 1.
- **Reticle**: represents the player’s weapon’s pointer in the screen. More important, weight: 0.5.

Suppose that we have 3 synchronization groups with different communication frequencies and score thresholds:

- **Degraded**: Less important entities; frequency: every 100ms.
- **Medium**: Semi important entities; frequency: every 50ms.
- **Optimal**: More important entities, frequency: every 10ms.

When the game starts and when any drastic change is noticed in the network traffic, entities are redistributed to the groups.

- **Without LoD**: All the entities will always be attributed to the optimal group. Their updates will be sent every 10ms and then will evenly compete for network resources, therefore will be impacted the same way by any network congestion. This configuration is not flexible and remain de same no matter the network capacity.
- **With LoD**: if there is no congestion on the network, all the entities will be on the optimal group. In case of congestion or low network capacity, the entities are redistributed to the groups using ducks as ER, and the formula in 4.4 to compute entities' current score coefficient. So if the groups thresholds are well set, the clouds will belong to the degraded group with updates sent every 100ms, the ducks will belong to the medium group with updates sent every 50ms, and the reticle will belong to the optimal group with updates sent every 10ms. This way the more important entities suffer less from the network congestion, reducing the overall traffic load. In case of drastic fluctuation in network congestion or capacity, the server automatically adjusts the traffic profile of the game by reassigning communication groups.

# 5. PILOT EXPERIMENT

The objective of this pilot experiment is to evaluate the impact of the LoD based adaptation in cloud gaming on the player's experience. In order to evaluate this impact, we observe and compare the reaction of players during a game session with and without the proposed approach.

## 5.1. My Duck Hunt video game

To conduct this experiment, the video game"My Duck Hunt" has been developed. It is a competitive shooting game inspired from the traditional Duck Hunt video game [25]. Here is the description of the game: Five kinds of entities evolve in the game scene: the reticle, the ducks, the flamingos, the gombas and the clouds. The player controls the reticle and should point the reticle on the target and shut to kill. The game is divided in 5 rounds or waves of ducks. The player has to achieve the following goals:

- Kill as many ducks as he can. For each duck killed, the player gains points.
- Do not kill flamingos.
- Protect flamingos from gombas by killing gombas. Each flamingo killed result in point loss.

The clouds are background decoration elements. The figure 7 shows a screen shot of the game. Here, the ducks are the entities with a black body and a green head. The flamingos are pink and the gombas are the brown entities on the floor.

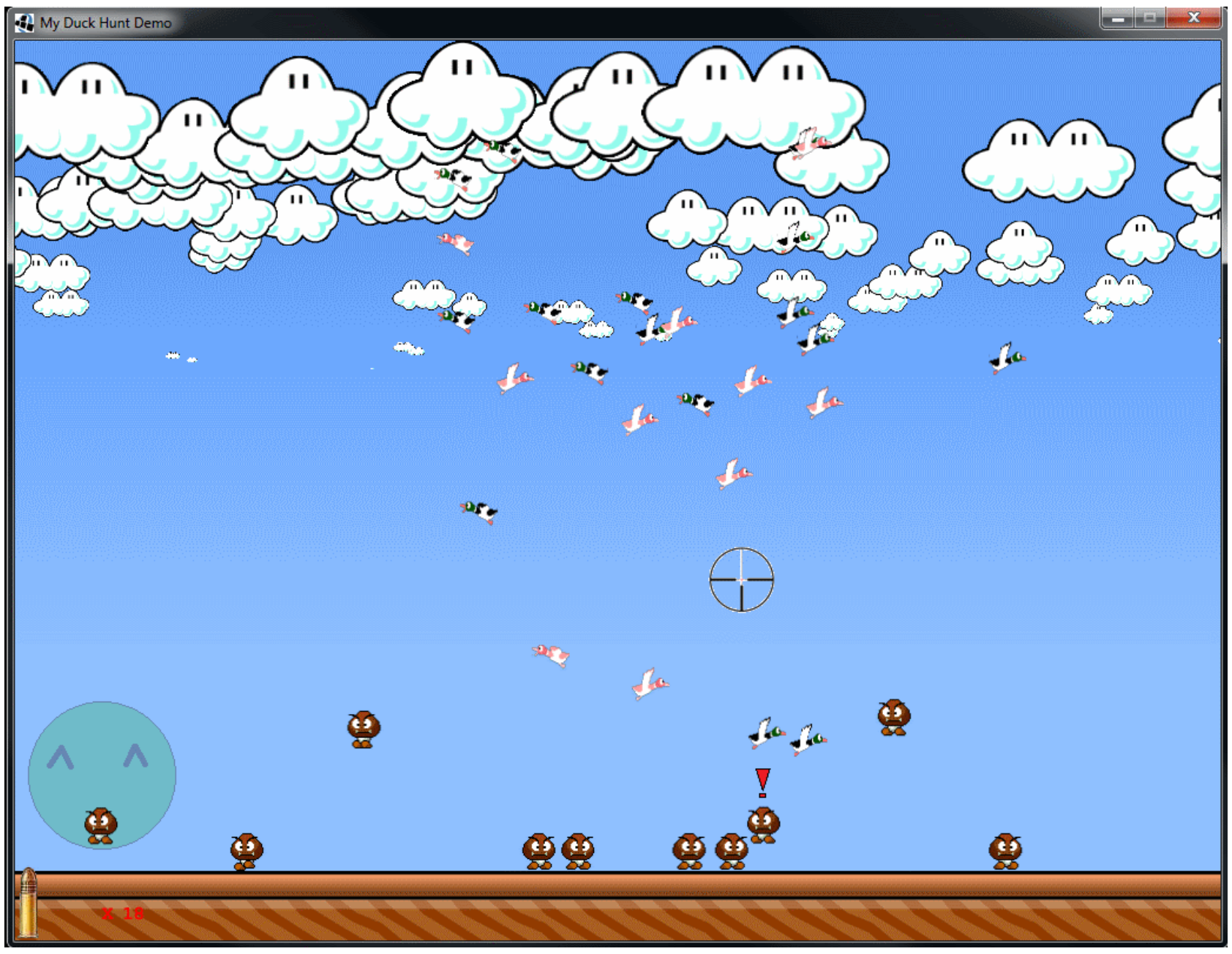

Figure 7: "My Duck Hunt" video game

## 5.2. Participants

The test was conducted with 9 participants between 21 and 30 years old with an average age of 25.33. The distribution of players, based on their playing frequencies; it is given in the table 1. Only one participant reported that he does not play video games. The other participants play games at least one time per week.

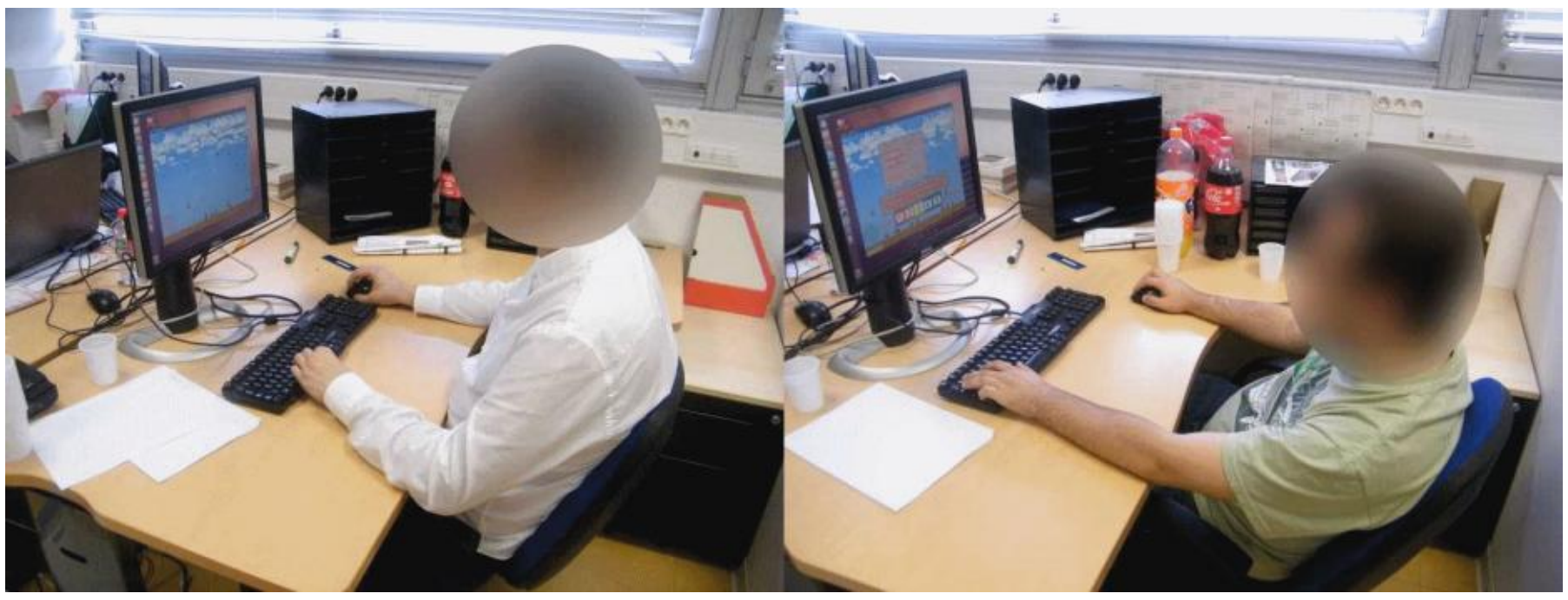

Figure 8: The candidates playing the game during the experiment

| **Never** | **1 per year** | **1 per month** | **1 per week** | **everyday** |
|---|---|---|---|---|
| 1 | 1 | 2 | 5 | 0 |

Table 1: Playing frequencies distribution

### 5.3. Protocol

The study follows a repeated-measures design[26]. The candidates have to play two versions of My Duck hunt. One that includes our LoD based proposition and the other without our proposition. In the later game, all the entities are synchronized at the same frequency. The experiment proceeds as follows:

- The candidates get a quick introduction of the game's rules through a demo version of the game.
- The candidates play one version then the other (this order is random for all the players). The candidates are not informed about the difference between the two versions. During the game, the candidates have to report when they perceive any type of bad quality of experience or interactivity shortage. They do so by holding the space key.
- At the end of each round, the candidates evaluate the quality of experience for the round. They give a mark between 1 and 5. 1 indicating a bad game experience and 5 indicating a good game experience.
- At the end of the experiment, the candidates are individually interviewed and asked to rate the global game experience for each game version by giving a mark between 1 and 5.

### 5.4. Network configuration

To be able to control the network environment, the pilot experiment is performed on a Local Area Network. In this LAN we have the game server machine, the client machine. Since we didn't deploy the server on a real cloud with WAN (Wide Area Network) connections, we need to simulate poor network settings of WAN. That is why we implemented a proxy.

The proxy is used for network congestion simulation through delay, jitter and packet loss simulation as illustrated in the figure 9. The proxy forwards all the packets from the client to the server and vice-versa. Since we are using UDP connections to send the state updates, the proxy simulates a congested network by ignoring all the packets received while a threshold of packets sent per second is reached. This threshold represents the capacity of the network or the available bandwidth: number of packets to forward per second. So to drastically change the network congestion for the game, we just need to change this threshold value.

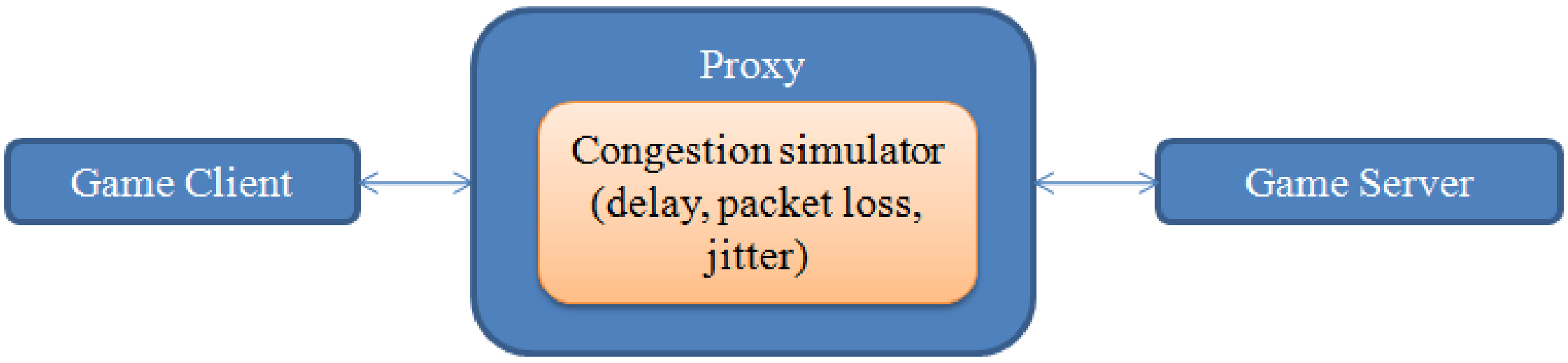

Figure 9: Congestion simulator via a proxy

The proxy is started at the same time as the game and it is launched with a configuration file dictating the variations in network capacity and therefore in network congestion during the game (jitter or unstable connectivity). The first 210 seconds of this configuration is given in the table 2. From 0 to 30 seconds, the proxy forwards 6000 packets per second in each direction; from 30s to 60s, it forwards 3000 packets per second in each direction, denoting a 50% network quality degradation, etc...

| **Time** | 0s | 30s | 60s | 90s | 120s | 150s | 180s | 210s |
|---|---|---|---|---|---|---|---|---|
| **Pkts/s** | 6000 | 3000 | 5000 | 2900 | 7000 | 2500 | 3500 | 3100 |

Table 2: Proxy configuration for network capacity

The game server is a Dell Precision M6500 with the following configuration: an Intel Core i7 Q 720 CPU and 4 Go of RAM. This is an experimental setup with a mono player game, meaning that the processing cost of the game can be handled by a server machine with these specifications. The only bottleneck we have is the one simulated by the proxy on the network link.

The server platform is configured with the four following communication groups sorted by other of importance:

- **Optimal group**: Entities with the highest communication requirements. In this group the update frequency of entities is 5 ms and the threshold score to stay in this group is 7.
- **Enhanced group**: Entities with relatively high communication requirements but lesser than those in the optimal group. Update frequency = 35 ms, threshold = 15.
- **Medium group**: Entities with average needs in network resources. Update frequency = 40 ms, threshold = 70.
- **Degraded group**: Entities with the lowest communication needs. Update frequency = 75ms.

## 5.5. Hypothesis

To evaluate the impact of the proposition we have stated the following hypotheses:

- **H.A.0** There is no difference in the game experience between the two versions of the game during round.
- **H.B.0** There is no difference concerning the global game experience between the two versions of the game.

- **H.C.0** There is no difference in the ratio of time spent holding the space key per the session duration between the two versions of the game.

### 5.6. Result

We use the paired t-test[27] to reject the three hypotheses. The statistical analysis was performed using R http://www.r-project.org version 2.15.0.
The hypothesis **H.A.0** is rejected for the five rounds with *p-value* < 0.5. The difference between the game experience during each round when using LoD and without LoD is statically significant. The results of this t-test are summarized in the table 3.

| **Round number** | **M** | **t(8)** | ***p-value*** |
|---|---|---|---|
| Round 1 | 1.2222 | 4.4 | 0.002287 |
| Round 2 | 1.555556 | 6.4236 | 0.0002039 |
| Round 3 | 1.666667 | 3.7796 | 0.005391 |
| Round 4 | 1.222222 | 3.0509 | 0.0158 |
| Round 5 | 1.666667 | 3.0151 | 0.01668 |

Table 3: Results

The hypothesis **H.B.0** is rejected by the t-test. The difference of the global game experience between the game with LoD based adaptation and the game without LoD is statically significant with a mean M = 1.777778, t(8)=8.6298, *p-value* = 2.521e – 05.

The hypothesis **H.C.0** is also rejected by the test. The different in the ratio of the time spent holding the space key per the game session duration between the two game versions is statically significant with a mean M = -0.1197444, t(8) =-2.4535 and *p-value* = 0.03972.

### 5.7. Discussion

The data gathered during the experimentation and the statistical study show the effect of our approach on the players' interactivity with the game, therefore their QoE. In fact all the three hypotheses were rejected meaning that the players have perceived a significant gain of the game experience with the adaptation technique not only for each round but also for the overall game session.

The t-test results have also rejected the hypothesis H.C.0, but with a higher *p-value* than the others. Because we choose a *p-value* threshold of 0.5, the hypothesis is still rejected but we can see that the results are not strong as in the two first hypothesis. A reason for this is that during the experiment, not all participants were following the guidelines we gave them about pushing the space key every time they notice a loss of interactivity or a decrease in the playability of the game, instead they were complaining verbally. Thus for some participants the data we gathered from the space key, does not reflect the quality of experience they actually experienced.

Finally, these results validate our approach on improving the overall quality of interactivity of the game, showing that adapting the traffic load generated by the game to the actual network capacity significantly increase the perceived quality of the game. Thus the approach keeps the game playable and enjoyable even in case of drastic variation in network congestion.

Our approach is not perfect, here are some limitations:

- Since the setting the configuration parameters (entity's weight, score coefficient threshold) of the platform is done manually by the game designer, the whole adaptation can be very subjective. A bad configuration of this system can result to a very bad player’s QoE. In our work, we suppose that the game designer knows what is doing (the significance of each entity) and the parameters are well set.

- It is also important to note that, while this approach reduces the bandwidth needed for a good quality game, it does not eliminate the requirement of a minimum bandwidth for an enjoyable game. It for sure requires less bandwidth than the classic cloud gaming services.

## 6. CONCLUSION

In this paper, we have proposed a LoD based network adaptation mechanism for cloud gaming. The main challenge in this context is the variability and sometimes the shortage of network connectivity which is required to enjoy a classic cloud gaming service. We address this accessibility and variability challenge by suggesting a new adaptation technique which uses entities organization model in order to:

- Minimize the effect of low and unstable network capabilities in maintaining game interactivity and improving player’s QoE.
- Take advantage of the accessibility, availability and scalability of the cloud computing in video games.

As future work, our next objective is to implement a version of the adaptation scheme using a topology heuristics (distance to camera) as the significance of the entities and perform a large scale experimentation in a multi-player cloud gaming environment where each player has his own camera and sees different angles of the game scene.

**Authors**

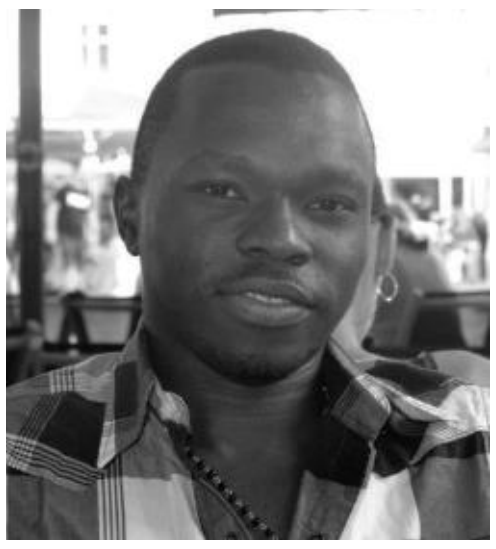

**Richard Ewelle** obtained his Master degree from the University of Lyon 1, Lyon, France in 2011, Bachelor of Computer science degree in 2008 from the University of Yaounde 1, Yaounde, Cameroon.

Currently, he is a Ph.D student in the Montpellier Laboratory of Informatics, Robotics and Microelectronics (LIRMM), Montpellier, France.

His specialization includes video games, network optimization and cloud computing.

**Yannick Francillette** is a Ph.D student in the Montpellier Laboratory of Informatics, Robotics and Microelectronics (LIRMM), Montpellier, France.

His specialization includes video games, game design, location based games and adaptation.

**Abdelkader Gouaïch** is an Associate Professor at the University of Montpellier, France.

His specialization includes video games, serious games, artificial intelligence and agent system.